\documentclass{sigchi}

\toappear{\scriptsize Permission to make digital or hard copies of part or all of this work for personal or classroom use is granted without fee provided that copies are not made or distributed for profit or commercial advantage and that copies bear this notice and the full citation on the first page. Copyrights for third-party components of this work must be honored. For all other uses, contact the Owner/Author(s). \\
{\emph{CHI '20, April 25--30, 2020, Honolulu, HI, USA.} } \\
\copyright~2020 Copyright is held by the owner/author(s). \\
ACM ISBN 978-1-4503-6708-0/20/04.\\
http://dx.doi.org/10.1145/3313831.3376656 }

\CopyrightYear{2020}
\doi{https://doi.org/10.1145/3313831.3376656}
\isbn{978-1-4503-6708-0/20/04}
\conferenceinfo{CHI'20,}{April  25--30, 2020, Honolulu, HI, USA}

\usepackage{balance}       
\usepackage{graphics}      
\usepackage[T1]{fontenc}   
\usepackage{txfonts}
\usepackage{mathptmx}
\usepackage[pdflang={en-US},pdftex]{hyperref}
\usepackage{color}
\usepackage{booktabs}
\usepackage{textcomp}
\usepackage{fancybox}
\usepackage{multirow}

\definecolor{postit}{RGB}{252,252,245}

\usepackage{microtype}        
\usepackage[all]{hypcap}    
\usepackage{ccicons}          
\usepackage[utf8]{inputenc} 

\def\plaintitle{Fairness and Decision-making in Collaborative Shift Scheduling Systems}
\def\plainauthor{Alarith Uhde, Nadine Schlicker, Dieter Wallach, Marc Hassenzahl}

\def\plainkeywords{Fairness; shift scheduling; healthcare; shift work; interview; nurse scheduling problem; hospital; conflict resolution; organizational justice; allocation norms; work-life balance; equality; need; equity; roster}

\makeatletter
\def\url@leostyle{%
  \@ifundefined{selectfont}{
    \def\UrlFont{\sf}
  }{
    \def\UrlFont{\small\bf\ttfamily}
  }}
\makeatother
\def\pprw{8.5in}
\def\pprh{11in}

\definecolor{linkColor}{RGB}{6,125,233}
\hypersetup{%
  pdftitle={\plaintitle},
  pdfauthor={\plainauthor},
  pdfkeywords={\plainkeywords},
  pdfdisplaydoctitle=true, 
  bookmarksnumbered,
  pdfstartview={FitH},
  colorlinks,
  citecolor=black,
  filecolor=black,
  linkcolor=black,
  urlcolor=linkColor,
  breaklinks=true,
  hypertexnames=false
}

\setcopyright{rightsretained}
\begin{document}

\title{\plaintitle}

\numberofauthors{4}

\author{%
  \alignauthor{Alarith Uhde\\
    \affaddr{Ubiquitous Design}\\
    \affaddr{Siegen University}\\
    \affaddr{Siegen, Germany}\\
    \email{alarith.uhde@uni-siegen.de}}\\
  \alignauthor{Nadine Schlicker\\
    \affaddr{Ergosign GmbH}\\
    \affaddr{Saarbrücken, Germany}\\
    \email{nadine.schlicker@ergosign.de}}\\
  \alignauthor{Dieter P. Wallach\\
   \affaddr{Ergosign GmbH}\\
    \affaddr{Saarbrücken, Germany}\\
    \email{dieter.wallach@ergosign.de}}\\
  \alignauthor{Marc Hassenzahl\\
    \affaddr{Ubiquitous Design}\\
    \affaddr{Siegen University}\\
    \affaddr{Siegen, Germany}\\
    \email{marc.hassenzahl@uni-siegen.de}}\\
}

\CopyrightYear{2020}
\setcopyright{rightsretained}

\maketitle

\begin{abstract}
The strains associated with shift work decrease healthcare workers' well-being. However, shift schedules adapted to their individual needs can partially mitigate these problems. From a computing perspective, shift scheduling was so far mainly treated as an optimization problem with little attention given to the preferences, thoughts, and feelings of the healthcare workers involved. In the present study, we explore fairness as a central, human-oriented attribute of shift schedules as well as the scheduling process. Three in-depth qualitative interviews and a validating vignette study revealed that while on an abstract level healthcare workers agree on equality as the guiding norm for a fair schedule, specific scheduling conflicts should foremost be resolved by negotiating the importance of individual needs. We discuss elements of organizational fairness, including transparency and team spirit. Finally, we present a sketch for fair scheduling systems, summarizing key findings for designers in a readily usable way.
\end{abstract}





\section{Introduction}
The shortage of skilled healthcare workers is a growing problem in many countries worldwide~\cite{Aluttis2014, OECD2007}. In an ageing society this becomes especially prevalent in geriatric care~\cite{Cooke2015}. Healthcare work is unappealing due to several factors, such as low pay, job-inherent emotional strain (e.g., confrontation with sorrow and death), and shift work~\cite{Cooke2015, Perrucci2007}.

Shift work creates particular difficulties for maintaining a fulfilling social life. For instance, long-term plans for private activities even on weekends often require careful arrangement ahead of time to secure the free days. Moreover, participation in social activities on a regular basis, such as a sports club, is more difficult than in common 9 to 5 jobs. The quality of the shift schedule in terms of both flexibility and reliability is crucial for the healthcare workers' ability to plan social activities and thus becomes central for their well-being.

However, creating a ``good'' schedule is difficult. First of all, legal regulations have to be taken care of. These include the maximum shift length allowed, the minimum breaks required between shifts, the qualifications as well as the extent of the jobs (full-time, part-time), and many more (see~\cite{VandenBergh2013} for an overview). On top of that, employees' preferences for certain shifts may be integrated. Taken together, these constraints result in a set of legal and economically feasible schedules, each having different consequences for the individuals concerned. When selecting one schedule from the set, these individual consequences need to be weighed against each other. Typically this selection is made by the supervising planner, who picks a schedule he or she considers as sufficiently ``fair'' for everyone.

Unlike the legal and economic criteria which can be formalized to a great extent, ``perceived fairness'' remains rather vague and subjective. The planner's understanding of what fairness means in the specific context may divert from the healthcare workers'. In order to maximize satisfaction with the shift schedule, we need a better understanding of this subjective fairness as the affected healthcare workers see it in order to implement truly ``fair'' algorithms. Therefore we set out to investigate healthcare workers' attitudes of subjective fairness during shift planning in the present paper. The objective is to better understand what determines the fairness of a shift schedule and how interactive systems can assist fair planning. To that end, we conducted two studies. In Study 1 we explore fairness in a detailed, qualitative analysis, scrutinizing the facets of fairness at the workplace and relevant resource allocation norms. In Study 2, the emerging themes are validated with a larger sample of healthcare workers from different institutions. Based on our results, we develop the sketch of a fair shift scheduling system for the healthcare domain.

\section{Previous Work}
\subsection{Hospital management}
Several support systems for various tasks are already available in the healthcare domain, including systems for task scheduling and documentation~\cite{Fitzpatrick2013}. One particular group are digital scheduling systems, which provide algorithmic support for solving shift planning problems (e.g.,~\cite{SiedaWeb, KronosWeb}). New algorithms have been developed to increase flexibility for the healthcare workers during planning and to better satisfy their preferences~\cite{Constantino2011, Constantino2015, VandenBergh2013, Lin1999, Lin2015}. However, this line of research approaches the shift scheduling problem mainly from a technical perspective and the healthcare workers' perspective is considered only secondarily.

The shift-related, HCI-driven research has mainly focused on shift handovers~\cite{Fitzpatrick2004, Tang2007, Zhou2010} or task scheduling in hospitals within a shift~\cite{Bardram2010, Bossen2016, Stisen2016}. The studied systems support the organization of work flows within and between hospital wards, such as patient, information, and equipment transfer~\cite{Bossen2016, Kuchera2011}. Another focus is the temporal organization of surgeries~\cite{Bardram2010, Egger1992, Kusunoki2015}. Shift planning itself has not been investigated extensively. One notable exception is a mobile application supporting self-care practices for shift workers~\cite{Nunes2018}. The app supports individual healthcare workers with establishing healthy chronobiological rhythms and with keeping track of their shift schedules. However, it does not address the actual creation of shift schedules.

\subsection{Fairness}
Recently, algorithmic fairness has gained some traction, with e.g., the newly founded specialized conference on Fairness, Accountability, and Transparency~\cite{FATWeb}. Besides many studies mainly considering technical implementations, some user studies have been carried out e.g., for computer-supported rent division~\cite{Gal2017, Goldman2015, Lee2017}, court decisions~\cite{Hou2017}, a dispatching system~\cite{Lee2017b}, and university course allocation~\cite{Othman2010}. However, with a few exceptions (\cite{Hou2017, Lee2017, Lee2017b}), these studies assume a model of ``objective fairness'', similar to the planning algorithms~\cite{Constantino2015, Lin2015}. This line of thought envisions a mathematically determined distribution of resources between groups or individuals, to provide ``provably fair'' solutions~\cite{Goldman2015}. However, different ``fair'' algorithms lead to conflicting distributions, which casts doubt on the claim of objectivity~\cite{Kleinberg2016}. Moreover, the decisions that these systems are ``fair'' are often made by the developers and can deviate from the users' judgements, leading to various problems when introduced~\cite{Binns2017, Selbst2019, Woodruff2018}.

Lee and colleagues~\cite{Lee2017, Lee2017b} compared different fairness concepts and distinguished between two fundamentally different \textit{allocation norms}: ``equality'' and ``equity''. The former implies that a resource should be distributed evenly among individuals, while the latter takes individual differences into account. Their definition of equity is based on Walster et al.~\cite{Walster1973}, who assume that all social interactions are fundamentally based on individual profit-seeking of the participants. In this line of thought, equity comprises all individual differences, including needs and performance. In contrast, Deutsch~\cite{Deutsch1975} argues that other motivations exist that justify to distinguish between needs and performance as separate norms. The goal of the cooperation determines which one is relevant: If the goal is economic productivity, he argues that performance (or ``equity'' in his definition) is most prevalent. However, if personal welfare is the primary goal, individual needs are dominant. In shift scheduling, for instance, equity could mean that people who work a lot of unpopular shifts such as Christmas Eve expect more flexibility in return (as a reward). In contrast, the need norm would provide more flexibility to e.g., single parents, who may have a hard time juggling children and a job, asserting their well-being. In this paper, we follow the more fine-grained definition by Deutsch with three separate allocation norms: Equality, equity, and need. Given their goal-dependence, different norms may be relevant within a group at different times~\cite{Deutsch1975}.

All of these different understandings of fairness mentioned so far are mostly considered with a fair distribution of resources. However, this is not the only aspect of fairness. In the organizational justice model broadly used in organizational psychology~\cite{Colquitt2001, Colquitt2015}, four facets of fairness have been researched (note that the terms ``fairness'' and ``justice'' are generally used interchangeably in the field~\cite{Colquitt2015}):

\begin{itemize}
    \item \textit{distributional justice} is the subjective fairness of the result of a decision-making process
    \item \textit{procedural justice} is the fairness of the decision-making process itself
    \item \textit{informational justice} describes whether the procedures are reasonable, complete, and on time
    \item \textit{interpersonal justice} is the interpersonal, respectful conduct
\end{itemize}

Previous research shows that all four facets are associated with job satisfaction~\cite{Susanj2012}. In the healthcare domain, Nelson and Tarpey~\cite{Nelson2010} found ``good'' shift scheduling and organizational justice to be positively related. HCI research has just recently begun to investigate the other facets, studying e.g., different explanation styles~\cite{Dodge2019}, decision-makers~\cite{Oetting2018}, and measures of objection~\cite{Lee2019}.

Current technical solutions (i.e., shift scheduling algorithms) neglect these facets of fairness and the majority is implicitly based on the equality norm~\cite{Constantino2015, VandenBergh2013, Lin2015}. In one case an equity norm~\cite{Lin1999} was applied. To our knowledge, no need-based computer-supported systems have been studied yet.

In the following, we explore healthcare workers' attitudes towards fairness in shift scheduling. Study 1 is an idiographic account of subjectively relevant fairness concepts, providing insights into which phenomena of the shift scheduling process are relevant for nurses and how they are interpreted in terms of fairness. Given the lack of HCI research on scheduling, it forms a starting point for further investigation. The analysis includes the facets of organizational justice and relevant allocation norms. Study 2 validates the emerging themes with a larger sample of healthcare workers from different institutions, allowing for causal inferences of the central findings from Study 1. We close with design implications for fair scheduling systems.

\section{Study 1: Interviews}
In Study 1, we explored the notion of subjective fairness in shift scheduling for healthcare workers. The specific research questions were:

\begin{itemize}
    \item Which allocation norms apply within the context of shift scheduling?
    \item How are the four facets of organizational justice perceived?
\end{itemize}

We used Interpretative Phenomenological Analysis (IPA)~\cite{Smith2009}, a method suitable for small samples that provides insights on sense-making processes and individual attitudes, matching our research goal (see e.g.,~\cite{Laschke2013, Rivituso2014} for other IPA studies in HCI).

\subsection{Participants}
Three female healthcare workers participated in our study (age: 24, 29, 57). Each held a degree as a registered nurse and had more than 6 years of working experience. The interviews took place in September 2017 during a morning shift and lasted around 35 minutes on average. Apart from the working time, they received no further compensation. In their organizational system, the nurses received their work schedule from the ward's planner, but one had an additional role as a deputy planner. They were recruited from different wards within the same retirement home and worked in different teams.

\subsection{Procedure}
The semi-structured interviews were based on an interview guide leading through questions about fairness from a general perspective to a more detailed description of specific experiences (see Figure~\ref{fig:interview_guide}). Three different interviewers accompanied their respective interviewee for two hours during their normal shift to establish mutual trust and familiarize with their environment before the interviews. We then ran the interviews in separate rooms in the residence, allowing for a calm, undisturbed atmosphere. All interviews were voice recorded. Given the open nature of the interviews, we sometimes deviated from the interview guide in order to gain a deeper understanding of the interviewees' perceptions and concerns. Afterwards, we debriefed the participants and thanked them for their time.

\setlength{\fboxsep}{5pt}
\begin{figure}[t]
\centering
\noindent\fcolorbox{black}{postit}{%
    \parbox{\columnwidth}{%
        \textbf{Interview guide}

        \vspace{5pt}

        What does fairness in general mean to you?

        How does a fair work-schedule look like for you?

        How would a fair process of work-scheduling look like?

        Which steps should be included in such a process?

        \vspace{5pt}

        Can you remember a situation in which you've felt that you were treated especially unfair/fair?

        How did you feel in this situation?

        How would you describe the difference between these two situations?

        \vspace{5pt}

        How do you think special contract conditions should be handled (e.g. someone works only in the early shifts)?

        How do you think these special conditions should be communicated?

        Do you think special contract conditions in general are fair?

        \vspace{5pt}

        Regarding fairness, which value does transparency hold for you?

        How would work-scheduling change if data about frequency of swapping or filling in would be collected?

        Who should have the permission to get insight into these data?

        \vspace{5pt}

        Regarding shift swap, are there some conflicts that come up frequently?

        Where lies the main problem in swapping shifts with other wards?

        Could you imagine working on other wards occasionally?
    }%
}
\caption{The interview guide for the IPA study.}~\label{fig:interview_guide}
\end{figure}

\subsection{Analysis}
Following the IPA guidelines~\cite{Smith2009}, we first transcribed all interviews in the native idiom of the interviewees as well as in standard German. The analysis was then conducted by the first and second author, independently. We started by simultaneously listening to and reading the transcript, while writing down first impressions. In the next step, everything considered noteworthy for answering the research questions was annotated regarding content, intonation, and expressed feelings. Then we summarized relations among concepts and emerging themes.

Finally, we compared and discussed the results of both analyses and consolidated or restructured themes accordingly, depending on agreement and disagreement. In the following section, we focus on shared themes, i.e., themes that emerged in at least two of the three interviews.

\subsection{Results}
We present the results structured by the four facets of organizational justice~\cite{Colquitt2001}, starting with distributional justice and the associated allocation norms. We then present procedural, interpersonal, and informational justice in greater detail.

\subsubsection{Distributional Justice and Allocation Norms}
Across all three interviews, a similar pattern of allocation norms emerged. With no further context given, all interviewees preferred shift scheduling based on equality. For one participant, fairness meant, \textit{``that everyone is treated the same way, no matter where they come from and what they do. Just equal rights''} [P3:19-20]. A general preference for equality in shift scheduling was even expressed unsolicited: \textit{``there are also situations [regarding the work-schedule] where employees get upset, when they see, for example... someone has only... one weekend off and someone else has three. That's a no-go and then they really get upset''} [P3:235-239]. Equality seemed to be the guiding allocation norm.

However, if more context was provided by talking through particular critical scheduling situations, the applied allocation norm changed. In response to one interviewee, who initially complained about a co-worker with a special agreement (an ``inequality'') to only work in the morning to be able to take care of children, we asked how these special agreements should be dealt with. Interestingly, she replied: \textit{``Well, actually it's fair, because they [employees with special contract conditions] also need their work...''} [P3:189-190]. Furthermore, she stated that in the case of conflicts among two healthcare workers the planner \textit{``needs to talk to the two parties. Maybe one only has an appointment with, let's say, the hairdresser and the other one has an important appointment with the doctor. In this case, I would say that the appointment with the doctor is more important''} [P3:47-50]. Both statements are clear references to the need norm. If equality were the norm, the nurse might have referred to respective preferences and an equal amount of granted wishes. Instead, she used the involved healthcare workers' needs as her criterion. A similar pattern emerged in the other interviews.

One problem that arises with need-based fairness is that employees, who don't have any obvious obligations beside their job, such as a family member to take care of, may experience a stronger pressure to be always available~\cite{Perrigino2018, Young1999}. We found this phenomenon, called \textit{family backlash}, among our interviewees as well. For instance, after being asked about a particularly unfair experience with scheduling, one participant commented: \textit{``Simply casual statements where someone said: `well, you are still young' and `You're still an apprentice with no other obligations' and `you could easily work 12 or 13 days in a row' [...] although I'm young, I still want my freedom and free time somehow. [...] Ok, so basically, for you I'm just a robot that needs to function and apparently I'm not allowed to have a private life [...] that was very stifling''} [P2:62-94]. In this excerpt, the nurse is especially upset about the assumption that she has no private obligations. She did not complain about the actual distribution of shifts, but about the fact that her availability is taken for granted, simply because she has no children or apparent health problems. Furthermore, it shows that the employee wants to have a say when it comes to decisions made about her work schedule. Her metaphor of feeling like a ``robot'' is on point: A robot has no needs or private life and can always work. It is a tool that gets told what to do and mindlessly executes its tasks. We understand this statement as a complaint that the nurse's personal needs are not appropriately taken into account.

For the sake of completeness, note that no statements supporting the equity norm were found in any of the interviews. All in all, while equality was mentioned as the overall relevant norm, specific situations were exclusively judged with reference to the need norm.

\subsubsection{Procedural Justice}
A predominant theme for the healthcare workers regarding the fairness of scheduling procedures was their involvement in decision-making. For instance, when talking about the distribution of shifts during public holidays, such as Christmas, one interviewee stated that \textit{``an optimal solution would be to discuss it within the team: `Who wants to work on which holiday?' ''} [P1:118-119]. Another interviewee highlighted that \textit{``he [the ward's planner] has always involved the employees while creating the work schedule: `Look at this, does it fit? Is it ok for you?' ''} [P2:151-153]. Yet another positive experience was: \textit{``[the planner] says at an early stage `Well, there are two shifts or two holidays and I cannot give everyone the day off'. You know, it became addressed as soon as possible and it can now be discussed within the team''} [P1:87-91]. From the deputy planner's perspective, it was mentioned that \textit{``when the work-schedule is finished, it's fixed and then we need to talk about it within the team and see whether there is a solution that is fine for everybody somehow''} [P3:246-248]. Notably, in these examples the healthcare workers are already more involved than just through the mere submission of preferences. However, this is not an explicit part of their current planning system, but a downstream informal practice based on individual planner's personal initiative. In fact, in the last excerpt, the interviewee explains that after the schedule is ``fixed'', employees engage in informal activities to cooperatively adapt it to everyone's needs, thereby actually circumventing the scheduling system and questioning the supposedly ``fixed'' schedule. Given that the healthcare workers mentioned the involvement as important for experiencing fairness, the active negotiation of the schedule should be considered to be made explicit within scheduling systems.

Not being involved leads to experiences of injustice. This became evident in statements such as: \textit{``Most of the employees are not even asked. They simply get their shifts assigned and only see it when the schedule is released''} [P2:478-480]. One participant explicitly complained about non-involvement: \textit{``But you need to ask me! You cannot just decide over my head''} [P2:410-411]. This experience can lead to negative emotions: \textit{``It happens very often that shifts are distributed without informing the employees. [...] And... well... I think it's sad''} [P2:166-168]. Both, the positive experience of involvement and the negative of non-involvement, are indicators in favor of a more participatory scheduling process, not only focused on the schedule as an outcome, but also the way it is created.

\subsubsection{Informational Justice}
The central topic concerning informational justice was the transparency of procedures. Its positive effect was described by one respondent: \textit{``I think it would be very important to have transparency here, because everyone could understand or accept why it [a scheduling preference] didn't work out this time. [...] And I think if the problems were openly addressed by the planner there wouldn't be so much resentment. Perhaps there would be more understanding''} [P1:340-355]. Accordingly, another participant pointed out the negative effect of intransparency: \textit{``Everybody wonders why the shifts are distributed and decided over our heads. [...] Why it is handled this way? Well, we don't know''} [P2:170-175]. Transparency is thus desirable insofar that it helps the healthcare workers to understand the reasons behind decisions important to them.

Conversely, however, the requirement of providing and justifying a preference publicly was seen as problematic as well: \textit{``I don't want to reveal to everybody why I want to swap''} [P1:402-403], and \textit{``Honestly, I believe nobody needs to justify why they need someone to replace them or why they are absent''} [P2:314-316]. If individual needs are the reason why a certain free shift cannot be granted, a context-sensitive compromise between privacy and transparency is needed that balances the requirements of everyone involved.

\subsubsection{Interpersonal Justice}
Concerning interpersonal justice, we found that the practice of asking a colleague to swap shifts requires an already existing, good and trusting relationship. The willingness of the person to cooperate is higher, the more they like the person who is asking: \textit{``A lot of it is handled by sympathy [...] if you know somebody privately [...] it's easier to ask or the willingness to swap is higher.''} [P1:395-400]. In a scheduling system based on cooperation and negotiation, positive relationships are a particularly important prerequisite.

In general, the respectful treatment of each other was crucial for experienced fairness and well-being. One interviewee described a hypothetical shift swap with a colleague: \textit{``If I had swapped shifts with a colleague in the past and now I want to swap one of my shifts [...] and he would reply: `No sorry, I don't feel like doing that.' [...] Well I mean, if he could give me a good reason or at least a reason at all, it's a different thing. But if it was like `No, forget it' directly... well then I would think it's unfair. And that would have been the last time for me to swap with him''} [P1:50-67]. This shows that the interaction of the co-workers is based on more than a mere quid pro quo social exchange. An overall cooperative attitude is crucial to successfully negotiate shifts. In this process, again, individual needs play an important role.

\subsection{Intermediate Summary and Discussion}
All in all, the interviews provided a detailed view on the prevailing concept of fairness among healthcare workers concerning shift scheduling. They cast doubt on a simple, equality-based distribution of shifts and suggest a stronger focus on the need norm. We found distributional justice to be a two-stage process, where equality is an underlying, rather abstract meta-norm. Nevertheless, when specific situations and conflicts are concerned, all interviewees preferred the need norm. Therefore, we assume that in the case of real conflicts or problems, an equality-based shift allocation may not be the most desirable choice. Instead, need-based conflict resolution seems more appropriate. Equity played no role in the interviews.

Scheduling procedures involving healthcare workers in the process were clearly preferred. Non-involvement was seen as unjust, independent from the question of whether the resulting distribution was perceived as fair.

Regarding informational justice, the two conflicting requirements of transparency and privacy were expressed. On the one hand, transparency is crucial to understand the decision-making procedure and the outcome. On the other hand, privacy is a requirement given that reasons for requesting time off can be sensitive. Mutual sympathy and trust is an important facilitator that can help to resolve this contradiction and a positive team spirit makes even personally critical decisions easier to accept. but a good solution for work places with a low social cohesion is still needed.

\section{Study 2: Experimental Vignette Study}
Our results from Study 1 were based on a small sample of three healthcare workers. While the in-depth interviews provided us with a very detailed understanding of their subjective fairness concepts, we ran an additional experimental study in order to test whether the central results are generalizable to a larger population of healthcare workers and whether differences in subjective fairness can be causally attributed to the different fairness norms.

We tested the following hypotheses concerning procedural and distributional justice in shift scheduling in the healthcare domain:

\begin{description}
    \item [H1:] On an abstract level, equality is more prevalent than the need (a) and equity (b) norms.
    \item [H2:] Specific scheduling resolutions are perceived as fairer when based on individual needs rather than on equality (a) or equity (b).
    \item [H3:] Specific scheduling resolutions are perceived as fairer when based on equality rather than on equity.
    \item [H4:] Specific scheduling resolutions are perceived as fairer when healthcare workers become involved in the decision-making process, compared with those made by a computer only.
\end{description}

H4 was not directly derived from Study 1, but added because we were interested in exploring the role of algorithmic systems for subjective fairness. In order to keep the number of variations low, we left the central, human planner out of our comparison of decision-makers, also because the difference between human and computer-made central decision-making has been studied before~\cite{Oetting2018}. Furthermore, we excluded aspects of interpersonal and informational justice in the experimental design. Both seem to depend to a great extent on the specific inter-individual relationship, which are difficult to address meaningfully in an experimental online study.

\subsection{Participants}
Participants were recruited in August 2018 through social media, snowball sampling, and from other residences of the same organization as the participants in Study 1. We only included healthcare workers in our study, who worked in shifts, were fluent in German, and who had no prior contact with us through our previous research. The participants could sign up for a raffle to win one out of five 30\texteuro{} vouchers for amazon Germany or Austria. Fifty-one healthcare workers participated in the study (13 male, 36 female, and 2 participants with undisclosed gender), with a median age of 35 years ($Min=21$, $Max=60$). Thirty-nine were certified nurses in general, geriatric, or childcare. Eleven were nursing assistants and one participant did not disclose his or her specialization. They were employed in hospitals (25), old people's homes (24), and a disabled people's home (1), with one undisclosed institution type. All participants worked shifts. Twenty had already been responsible for shift planning to some extent. The majority lived in Germany (48) and three participants lived in Austria.

\subsection{Procedure}
Completing the entire study took approximately 15 minutes. After accessing the introduction page through a link, the participants were first informed that the topic was shift scheduling, data collection was anonymous, and that they could stop at any time. Next, the participants filled in demographic questions asking about their specialization, age, gender, and country of residence. Here we filtered out participants not working shifts and/or not working in healthcare. Following the demographics, each participant was asked to formulate in two open questions what fairness means to them in general and with regard to shift scheduling. We chose this open format to avoid directing the answers towards a specific norm.

Participants were then randomly assigned a set of 18 \textit{vignettes}. Each vignette represented a brief description of particular situations in shift scheduling. Systematically varied vignettes are at the heart of \textit{Experimental Vignette Methodology} (EVM)~\cite{Aguinis2014}, a method frequently used in Social Psychology and its applied fields (e.g.,~\cite{Tversky1981}). Typically, participants are asked to imagine the presented situations and to make explicit decisions, judgments, or to express behavioral preferences within these situations. This method allows for experimental rigidity in realistic, contextualized scenarios.

In the present study, the vignettes described a conflict between two healthcare workers. Both wanted a day off -- the participant and a (fictional) co-worker to whom they had a neutral relationship. An example vignette was: \textit{``It's you or your co-worker -- one of you has to work. You want your day off because \textbf{you have an important appointment with your doctor} [need norm salient]. On the other hand, it's your co-worker's turn, because \textbf{he got almost none of his wishes granted recently} [equality norm salient]. \textbf{The system decides} [locus of decision] in favor of \textbf{your co-worker} [winner], in order to assure that the free time is distributed equally among all employees. [reason]''}.

We varied the \textit{norms} that the participant and co-worker used to justify their respective claims for the free shift. Furthermore, we varied the \textit{locus of decision} (healthcare workers vs. system) and the \textit{winner} (who gets the free shift?). The vignettes ended with one sentence explaining why the decision was made to clarify the applied allocation norm. Thus, the fixed context factors we provided were the neutral relationship with the co-worker (1), the need to decide for one of the two shifts because of staff shortage (2), and the medium level of transparency (3): The provided reason stated the allocation norm of the winner, but no further details. See Table~\ref{tab:vignettes} for all variations used.

These variations resulted in a total of 36 vignettes (3x3x2x2). We chose a set size of 18 vignettes for each participant, allowing us to test our hypotheses without confounding factors at a reasonable study duration~\cite{Atzmueller2010, Steiner2006}. We varied the factor \textit{locus of decision} between individuals, while all other variables could be tested as repeated measures. This means that, for example, an individual participant only answered vignettes in which the computer made decisions, and did not see those where the co-workers decided together (and vice versa). However, each participant saw several scenarios with different argumentations based on needs, equality, and equity, in which sometimes the participant won and sometimes the co-worker. In addition, we used two different operationalizations for each argument norm in order to increase variation and to allow for more natural comparisons when both healthcare workers in the vignette justified their claim with the same norm (e.g., need vs. need). Then, two sets of 18 vignettes each were created separately for both factor levels of \textit{locus of decision} (healthcare workers, system), with randomized operationalizations and orders of the vignettes. Thus, we had a total of four sets covering all possible combinations. For each vignette, the participants rated the fairness of the result (distributional justice) and the fairness of the process (procedural justice) on a 7-point scale ranging from ``unfair'' to ``fair'', together forming a comprehensive measure of the tested subjective fairness facets. In addition, we asked how well they could imagine that situation (7-point from ``not well'' to ``well''). The study concluded with a comment field and the possibility to leave a contact address for the raffle.

\begin{table*}[t]
  \centering
\begin{tabular}{lll}
    \toprule
    \multicolumn{3}{c}{\textbf{Vignette Variations}}         \\
    \midrule
    \multirow{6}{*}{Argument Norms} & need & ``because you have/your co-worker has an important appointment with the doctor.'' \\
             & & ``because you/your co-worker want(s) to go to a friend's wedding.'' \\
    & equality & ``because you have/your co-worker has gotten almost no wishes granted recently.'' \\
            & & ``because you/your co-worker has had almost no weekends off recently.'' \\
    & equity   & ``because you/your co-worker stand(s) in for sick co-workers all the time.'' \\
           &  & ``because you/your co-worker had especially exhausting shifts recently.'' \\[6pt]
    \multirow{3}{*}{Locus of Decision} & system & ``the system decides in favor of ...'' \\
    & nurses   & ``the system asks you and your co-worker to find a solution. Together, you decide \\
    &          & in favor of ...'' \\[6pt]
    \multirow{2}{*}{Winner} & participant & ``yourself''                                    \\
   & co-worker & ``your co-worker''                         \\[6pt]
    \multirow{3}{*}{Reason}  & need   & ``because your/your co-worker's need seems to be higher at that time.'' \\
      & equality & ``in order to assure that the free time is distributed equally among all employees.'' \\
      & equity   & ``as a recognition for your/your co-worker's commitment.'' \\[6pt]
    \multirow{2}{*}{Justice Facet} & Distributional Justice & ``I think the result is fair''\\
                                   & Procedural Justice     & ``I think the decision-making process is fair'' \\
    \bottomrule
\end{tabular}
  \caption{The variations used in the vignettes. ``Reason'' varied based on the two arguments and the winner, so it was not an independent variable.}~\label{tab:vignettes}
\end{table*}

\subsection{Results}
\begin{figure}[t]
    \centering
    \includegraphics[width=\columnwidth]{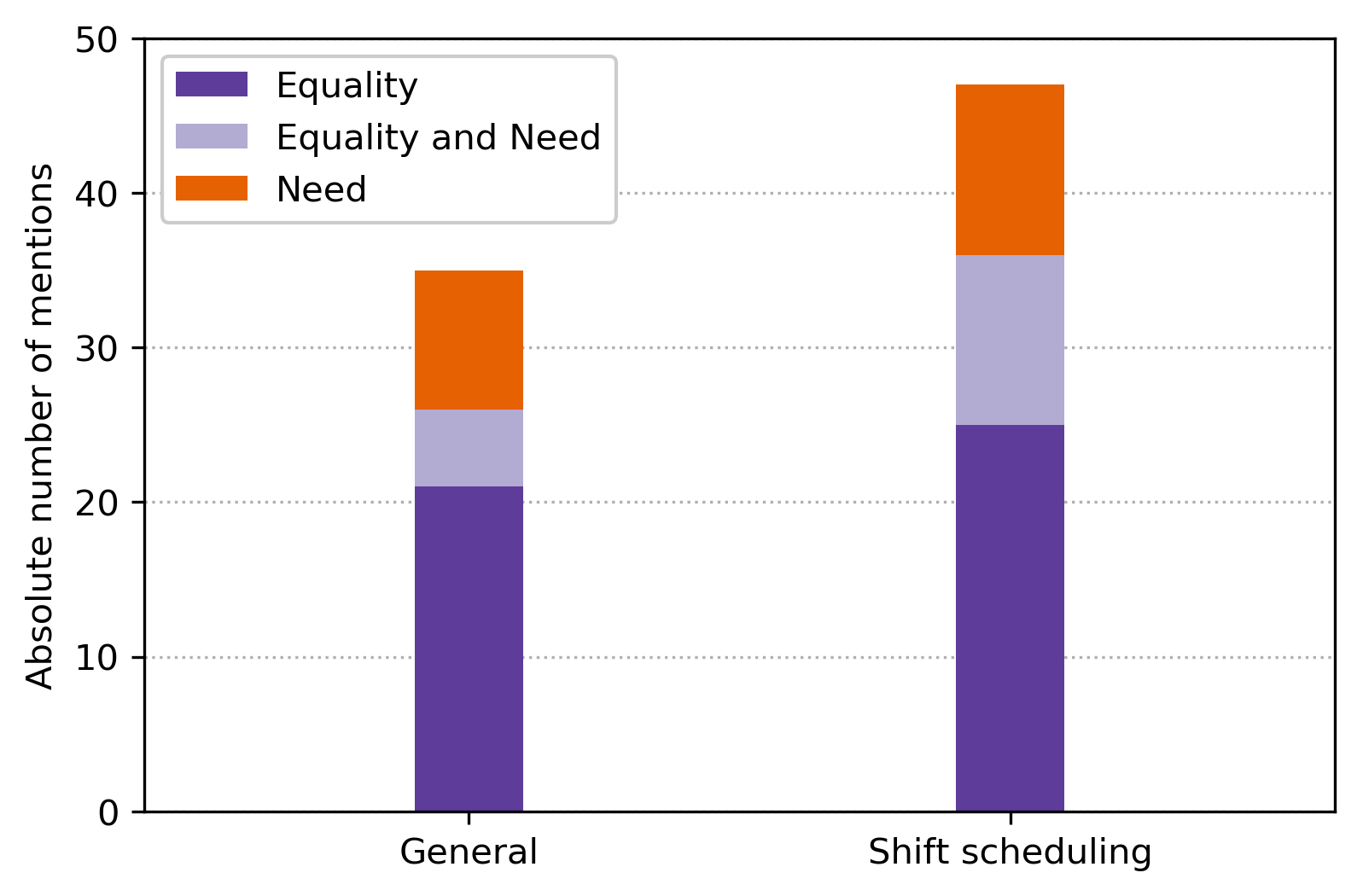}
    \caption{Absolute frequencies of allocation norms the healthcare workers mentioned when asked about fairness in general (left) and in shift scheduling (right). Equity was not mentioned. (n=51) (CC BY 4.0 \ccby)}~\label{fig:frequencies}
\end{figure}

First, we analyzed the open-ended questions, where participants described what fairness means to them, both in general and in relation to shift scheduling. Two independent coders (the first and second author) categorized each answer with respect to the implied allocation norms: need, equality, and equity. Multiple categorization was allowed. If a clear categorization was impossible, the statement was coded as \textit{no association}. For example, one participant wrote: \textit{``Justice for all''}, which could not clearly be attributed to any category. On the other hand, \textit{``Considering the needs of every human being''} was assigned to the ``need'' category and \textit{``That everyone is treated in the same way and nobody receives preferential treatment''} was assigned to ``equality''. The inter-rater reliability across both questions was very good ($\text{Krippendorff's }\alpha=.86$,~\cite{Hayes2007}) and remaining disagreements were resolved in a follow-up discussion. The absolute frequencies of mentioned norms and their combinations are shown in Figure~\ref{fig:frequencies}.

When asked about fairness in general, 26 of the 51 participants referred only to the equality norm ($41\%$; $CI(.95)=[28\%, 56\%]$), followed by 9 referring only to the need norm ($18\%$; $CI(.95)=[9\%, 31\%]$). Moreover, 5 comments were assigned to both equality and need ($10\%$; $CI(.95)=[4\%, 22\%]$). No participant referred to equity and 16 comments could not be categorized. When asked about fairness in the context of shift scheduling, 25 participants ($49\%$; $CI(.95)=[35\%, 63\%]$) related their answer to equality, 11 ($22\%$; $CI(.95)=[12\%, 36\%]$) to needs, and 11 to both equality and needs ($22\%$; $CI(.95)=[12\%, 36\%]$). Again, nobody referred to the equity norm. Four comments could not be categorized. Besides these three categories, a few participants also mentioned aspects relating to interpersonal and informational justice. For example, one participant mentioned \textit{``friendly cooperation''} among other aspects and another one wrote: \textit{``Honest, open, transparent, just''}.

As expected (H1), equality was the participants' primary fairness norm in general (41\% exclusively + 18\% together with needs = 59\%) and when specifically, but abstractly asked about shift scheduling (49\% exclusively + 22\% combined with needs = 71\%).

\subsubsection{Vignettes}
To analyze the vignettes, we first transformed the data into the "tidy data" format~\cite{Wickham2014}, removing redundant coding. Given that the factor levels of both healthcare workers' \textit{norms} were equivalent, we could reduce our model to a 2x3x2x2-MANOVA without losing information, with the within-subject factors \textit{argument role} (winning vs. losing \textit{argument}), \textit{argument norm} (need, equality, and equity), and \textit{winner} (participant vs. co-worker), as well as the between-subjects factor \textit{locus of decision} (healthcare workers vs. system). For example, when the participant won with a need argument against an equity argument by the co-worker, we coded the winning argument as need-based, the losing argument as equity-based, and the participant as the winner. Our dependent variable was a combination of the two \textit{justice facets} (distributional justice and procedural justice) in a single, compound measure. We had no specific hypotheses for the individual facets.

Imagining the situations was possible with a high average of 5.57 points ($SD=1.37$) on a scale from 1 to 7. Thus, the vignettes represented realistic situations our participants could relate to.

The MANOVA revealed a significant interaction effect for \textit{argument role} x \textit{argument norm} using Wilks's statistic ($\Lambda=.63$, $F(4, 194)=12.75$, $p<.01$, $\eta_p^2=.21$). Moreover, we found a main effect of \textit{locus of decision} on fairness ($F(1, 49)=3.31$, $p<.05$, $\eta_p^2=.06$). Most other interaction effects were confounded with the set effect, because their factor levels could not be equally distributed among them~\cite{Atzmueller2010} and they were also not significant. The two non-confounded effects were the two-way interactions that included the \textit{argument norm} but not \textit{argument role} and were thus difficult to interpret (and also non-significant). The remaining, interpretable main effect of \textit{winner} was also not significant ($F(1, 49)=1.50$, $p=.49$, $\eta_p^2=.01$). Both significant effects will be analyzed in more detail in the following.

\begin{figure}[t]
    \centering
    \includegraphics[width=\columnwidth]{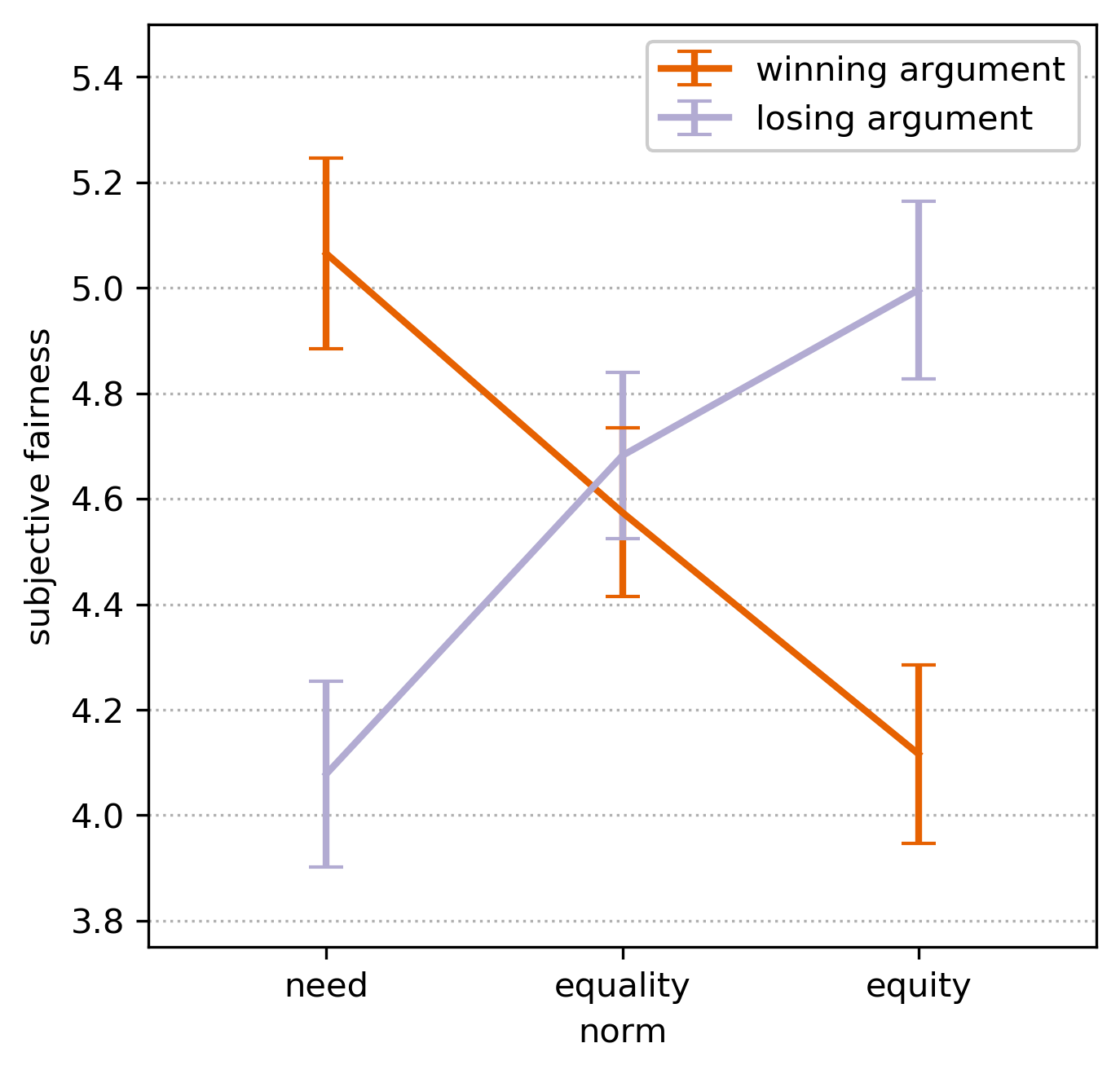}
    \caption{The interaction effect \textit{argument role x norm} from the vignette study. Y-Values represent the combined justice facets. Error bars represent the standard error. (n=51) (CC BY 4.0 \ccby)}~\label{fig:anova1}
\end{figure}

The interaction effect relates to our hypotheses H2 and H3. As expected, decisions based on a need argument ($M=5.06$, $SE=0.18$) were perceived as fairer than those based on an equality argument ($M=4.57$, $SE=0.17$, $t(50)=3.00$, $p<.01$, $d=0.42$) and fairer than decisions based on an equity argument ($M=4.11$, $SE=0.18$, $t(50)=5.07$, $p<.01$, $d=0.71$). Moreover, when a decision was based on an equality argument it was perceived as fairer than one based on an equity argument ($t(50)=3.08$, $p<.01$, $d=0.43$). Conversely, when a need argument was ignored ($M=4.07$, $SE=0.18$), the decision was perceived as less fair than when an equality argument was ignored ($M=4.68$, $SE=0.16$, $t(50)=4.14$, $p<.01$, $d=0.58$) or when an equity argument was ignored ($M=4.99$, $SE=0.17$, $t(50)=4.95$, $p<.01$, $d=0.69$). Decisions ignoring equality arguments were perceived as less fair than those ignoring equity arguments ($t(50)=2.30$, $p<.05$, $d=0.32$). The results are depicted in Figure~\ref{fig:anova1}. In sum, decisions based on the need norm were perceived as most fair and those ignoring the healthcare workers' needs were perceived as least fair, confirming H2. Moreover, decisions based on the equality norm were perceived as fairer than those based on the equity norm and decisions against the equality norm were less fair than those against the equity norm, confirming H3.

The main effect of \textit{locus of decision} supports our exploratory hypothesis H4. Those cases in which the healthcare workers made a collaborative decision had a higher subjective fairness ($M=4.84$, $SE=0.20$) than those in which the system made a decision autonomously ($M=4.33$, $SE=0.20$).

\subsubsection{Open comments}
Finally, we analyzed the open comments to see whether they further qualify our results. In line with H4, two participants wrote that they see mutual decision-making between two employees as preferable to computer decisions. One participant stressed the fairness of mutual agreement: \textit{``If the system asks us to find a decision together‚ and we do that, then the decision who gets the time off is based on mutual agreement. So, logically, it is fair''}. Another one focused on problems with computer decisions: \textit{``I don't like the idea to have an algorithm decide about my free time''}.

Two more comments provide suggestions on how to act if no mutually acceptable decision can be found based on the need norm. One person commented: \textit{``If you have a co-worker who always insists on his free time, no matter how important your appointment is, I would draw lots. Thank god that hasn't happened yet''}; and the other one: \textit{``I would accept generally unique reasons like a wedding for a day off, as long as it's not the wedding of a person the employee hardly even knows. [...] Everyone has the right to have some days off, but also has to accept if it doesn't work out sometimes. However, if it turns out that someone is not reliable or `often sick', maybe an employee who is more reliable should be preferred''}. In both cases, the need norm has the highest priority. But if it is not sufficient or misused, one healthcare worker suggests a random decision while the other one prefers a performance-based process.

\subsubsection{Subjective fairness in case of conflicts}
In sum, when looking at fairness in general, the equality norm was most common, confirming our hypothesis H1. However, specific conflict resolutions based on the need norm were seen as the fairest, followed by the equality norm, as expected with H2 and H3. Finally, decision-making that was guided by the computer but made by the healthcare workers was fairer than autonomous decision-making by the computer, confirming H4.

\section{Summary and Discussion}
Three in-depth interviews and an experimental vignette study helped us draw a detailed picture of subjective fairness in shift scheduling. The oversimplified concept of fairness as equality of outcomes underlying the majority of shift scheduling systems (e.g.,~\cite{Constantino2015, Lin2015}), does not capture healthcare workers' understanding of fairness. While equality was the preferred norm on an abstract level, it is not deemed appropriate for resolving concrete scheduling conflicts. Instead, conflict resolution should be need-based. Equity, that is, fairness based on prior performance, did not play a pronounced role. It was also considered as the subjectively unfairest norm in our experimental study, while the need norm was the fairest, which supports our distinction following~\cite{Deutsch1975} and contrasting previous work~\cite{Lee2017, Lee2017b, Walster1973}. Moreover, our exploratory comparison with H4 shows that previous systems that are designed to make autonomous decisions about personally relevant conflicts may profit from integrating the healthcare workers in the process. The scheduling algorithm should ideally support the participants in their decision-making, but not decide on their behalf. Notably, we found this effect despite the fact that both the algorithm-based and collaborative decisions were briefly explained to the healthcare workers in each vignette, thus providing higher explainability than existing systems~\cite{Constantino2015, Dodge2019}.

Why do current solutions tend to apply the equality norm? We suspect that one reason is the dream of a fully automated scheduling process that solves the social conflicts autonomously. This requires an exhaustive mathematical model of fairness. When creating such a model based on needs, one requires a vast range of (partially sensitive) information about each worker in each situation. For instance, assume a conflict where one participant ``takes care of the own child'' which could be considered an important reason, while the other one ``goes to a concert'', which might be less important. However, the urgency of the former changes depending on whether or not one has e.g., a spouse to help out, compared to being a single parent. In contrast, the latter may be subjectively more important if it is part of a romantic date and as such possibly a step on the way to build a family. A fully automated system needs these sensitive data to make meaningful decisions, which remain normative. In contrast, equality as used so far (e.g.,~\cite{Constantino2015}) is less (but still) complicated to implement. Moreover, the qualitative data from Study 1 indicate that shift scheduling may be a case of ``particularized justice''~\cite{Binns2020}, meaning that it is important to treat each conflict on a case-by-case basis and not as another iteration of learned, appropriate decisions. This implies fundamental issues when treating it as a purely mathematical decision problem~\cite{Binns2020}. Instead of aiming for full automation, we thus advocate a meaningful integration of the affected workers, not as an intermediate step, but as a design goal. Leaving the decision to the healthcare workers then seems to be a logical and straightforward solution that can be supported by technology. We showed that such a mutual decision-making process is preferred by the healthcare workers over a fully autonomous algorithm.

\subsection{Design implications: A sketch of a fair scheduling system}
\begin{figure}[t]
    \centering
    \includegraphics[width=\columnwidth]{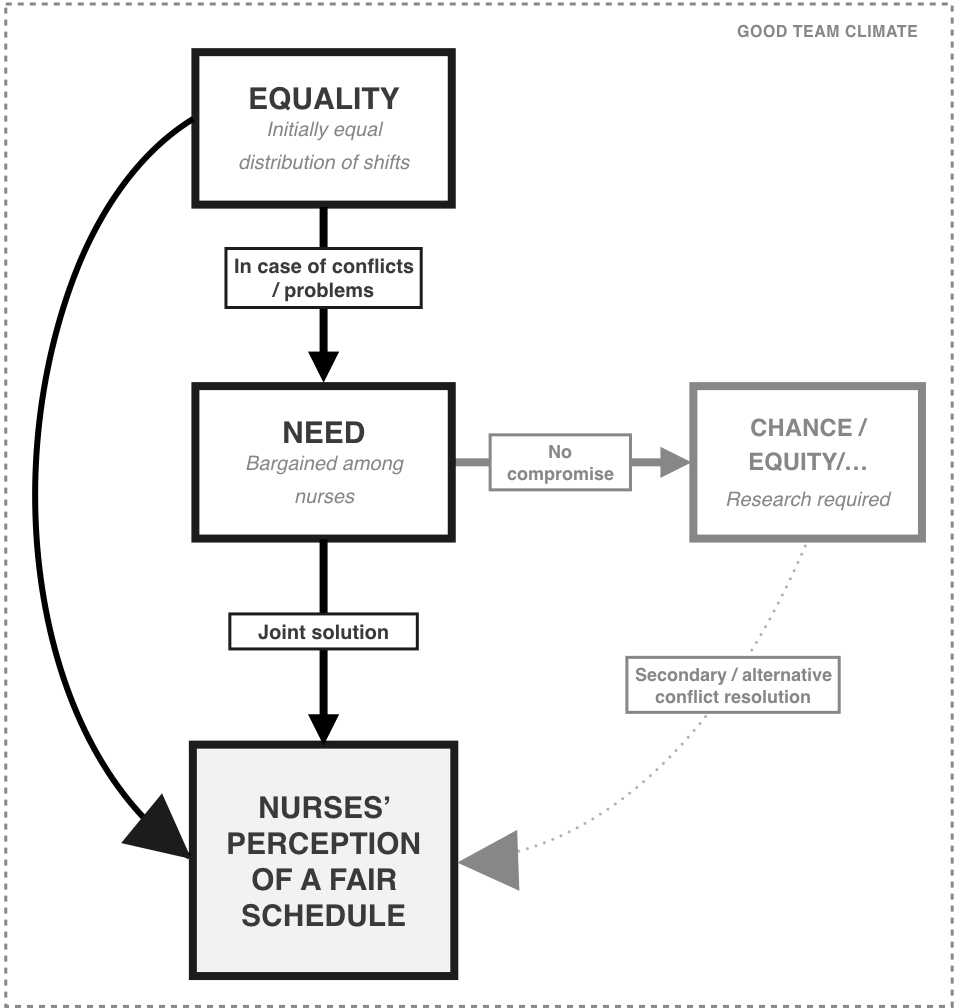}
    \caption{A sketch of a fair shift scheduling system. On a general level, equality should be the goal. However, when conflicts arise, they should be solved on a need basis. If this doesn't lead to a solution, other processes may be used. Generally, a good team climate facilitates the negotiation process. (CC BY 4.0 \ccby)}~\label{fig:model}
\end{figure}

Based on our findings, we are able to outline a scheduling system designed to increase subjective fairness in shift planning for healthcare workers (see Figure~\ref{fig:model}).

\subsubsection{Equality in general scheduling decisions}
On an abstract level, all healthcare workers should be treated equally. This includes that they should have similar amounts of wishes and free weekends, and similar abilities to include their preferences in the schedule. Algorithms are good at asserting this equality.

\subsubsection{Need-based conflict resolution}
However, in case of a specific conflict due to overlapping wishes, healthcare workers must be given the chance to justify their claims by stating why they need the time off. These needs should form the basis for conflict resolution.

\subsubsection{Inclusive decision-making}
The conflict resolution should not be done by the system itself, since this leads to lower procedural justice than mutual decision-making. Instead, the system should support conflict resolution by fostering a sensible, step-wise procedure of disclosing the claims to each other, with a focus on building mutual trust and maintaining privacy. Computer-support could be included in the form of finding the scheduling conflicts, presenting all legal solutions (e.g., who would have to withdraw a preference), and indicating how and when the healthcare workers could resolve the conflict in advance, e.g., the next time they work together. A system like this would turn a now informal practice heavily relying on a committed planner to a more formal practice open to all healthcare workers in an organization.

\subsubsection{Two options for difficult conflicts: Randomness or equity}
Probably not every conflict can be resolved with mutual agreement. In such a case, either an equality-based mechanism, such as granting a similar number of ``wins'' to each healthcare worker, or an equity-based solution, such as rewarding flexibility, may be used. However, while these mechanisms seem straightforward at first, on closer inspection a number of issues arise. For instance, one could use the overall number of granted wishes for each healthcare worker (with and without conflict) or focus on the number of ``wins'' and ``losses'' of the healthcare workers involved in the conflict at hand only. Moreover, the time span has to be taken into account: how much does a ``loss'' from three months ago count in comparison to one from five years ago? In addition, an employee with a part-time position may be more flexible to help out than the full-time employee who is at work anyway. When applying an equity-based system, how do we compare the demonstrated flexibility these two healthcare workers show? While equity and equality seem to imply a certain objectivity, there is no natural ``true'' standard for all these decisions. Needs can be expressed and negotiated, and in case no agreement can be found, a random selection (i.e., flipping the coin) is a simple solution avoiding all these complexities. Alternatively, the specific team could agree on the exact conditions of how equality and equity is valued or even attempt to avoid a forced decision altogether by promoting consensus-orientation (see e.g.,~\cite{Krawinkler2018, Laloux2014}). In a future study, we plan to investigate the respective experiential implications further.

\subsubsection{Cultivate a positive team spirit}
Respectful conduct is an important prerequisite for successful shift planning. In Study 1 we found that mutual respect and empathy for the co-worker increased the willingness to swap shifts, i.e., to engage in constructive conflict resolution. A shift scheduling system should therefore foster this positive team climate, e.g., by explicitly designing for positive interactions among the healthcare workers through the system. This implies, for example, the design for different levels of informal communication.

\subsubsection{Data souvereignity and transparency}
To successfully resolve conflicts, the system needs to balance privacy and transparency. On the one hand, some healthcare workers don't want to share private, sensitive information publicly. On the other hand, knowing the reason why a free shift is granted to a co-worker helps accepting the decision (if the reason is deemed valid). The system should support the negotiation by initiating it and by facilitating the stepwise, voluntary disclosure of private information until a solution is found. The disclosure can also happen offline, e.g., in a private conversation at work.

In sum: Despite potential concerns about ``automated'' decision-making, algorithmic scheduling systems can be of help if designed correctly. They can create legal schedules and detect scheduling conflicts among healthcare workers. Since not all conflicts are evitable, it can facilitate the resolution by fostering need-based negotiations. The system can jump-start constructive discussions among involved healthcare workers and, for instance, inform them when they share a shift to talk solutions through face-to-face. It can also show possible solutions so that healthcare workers can be sure that their rescheduling (e.g., of a doctor's appointment) actually solves the problem. This facilitates pro-social behavior and may be beneficial for the team climate, fueling a virtuous circle of growing together as a team. Finally, by indicating and explaining legal issues, a scheduling system allows healthcare workers without further training to engage in scheduling and take back control of their time.

\subsection{Limitations}
While we believe that our suggested approach to shift scheduling would have a positive impact on the experience of fairness and well-being of healthcare workers, there are a number of possible negative consequences to be considered~\cite{Hecht2018}. First, a need-based system risks to further consolidate the so-called ``family backlash''~\cite{Perrigino2018, Young1999} by implicitly giving an advantage to healthcare workers with higher needs due to family obligations. However, this risk might be less severe than it appears at first glance. Study 1 showed that healthcare workers already maintain practices of adapting presumably ``fixed'' schedules to their needs. The suggested system would support a better match between the ``official'' and the emerging ``real'' shift plan. A possible family bias could therefore be more accurately represented, providing explicit data that may help to develop coping mechanisms within the team. In addition, even the young apprentice nurse from Study 1, who has no own children, found need-based scheduling to be fair in principle. When introducing an explicitly need-based conflict resolution process, needs in general may become more visible, since it becomes official practice to base decisions on individual needs and negotiations. This may in turn even reduce the bias. However, without an implemented system, this remains speculation.

Second, the system outlined assigns additional tasks to the healthcare workers, namely conflict resolution and deeper involvement in planning. This adds to their already high work load. While we cannot rule out the possibility that some healthcare workers will find this problematic, the interviewees in Study 1 mentioned a few situations where they would prefer a stronger involvement. In fact, there were no complaints about too much involvement in either study. This is consistent with previous insights into organizational justice, where more involvement is generally perceived as positive~\cite{Colquitt2015, Nelson2010} despite the added work. Moreover, as we found in Study 1, healthcare workers already involve themselves into shift scheduling to varying degrees, currently as an informal practice. Through the suggested need-based negotiations, these practices would become an official part of the everyday job responsibilities, which may actually make already existing efforts more visible.

Third, results in Study 1 were based on a small sample. We replicated the central findings in Study 2, but more confirmatory research is needed e.g., for interpersonal and informational justice. The existing data indicated that both team coherence and a balance between privacy and transparency are important.

\section{Conclusion}
Creating shift schedules is complex. While algorithms could and should support this task, it seems important to not only consider the objective quality of the resulting shift schedule (i.e., from an economic or legal point of view), but also the subjective experiences of the people involved. Employees' well-being in shift work is closely intertwined with the way their shift schedule is determined. We identified perceived fairness as a crucial aspect and sketched an inclusive system with fairness as a primary concern. Our study dismisses overly simplistic ``one size fits all'' notions of fairness. It draws a granular picture of interrelated notions of distributional, procedural, informational, and interpersonal justice, currently often expressed and fulfilled in informal practices outside the official procedure for scheduling. We believe it to be important to better understand these practices and to make them ``official'' by explicitly addressing them in the design of (semi-)automated, cooperative scheduling systems. In this sense, we understand the focus on fairness of the work schedule as a contribution to foster well-being and job satisfaction in healthcare.

\section{Acknowledgments}
The project is financed with funding provided by the German Federal Ministry of Education and Research and the European Social Fund under the ``Future of work'' programme (grant number 02L15A216).

We would like to thank Christina Krahl for her help with the interviews.


%
%
%
%
%

\balance{}

\bibliographystyle{SIGCHI-Reference-Format}
\bibliography{/home/alarith/Dokumente/bibliography}

\end{document}